\documentstyle[11pt]{article}
\newcommand{\bra}[1]{\langle#1|}
\newcommand{\ket}[1]{|#1\rangle}

\newcommand{\La}{\Lambda}
\newcommand{\la}{\lambda}
\newcommand{\slsh}[1]{#1\hspace{-0,22cm}\slash}

\title{ \bf{4$^{th}$ order similarity renormalization \\
	of a model hamiltonian}}
\author{Tomasz Mas{\l}owski and Marek Wi\c eckowski}
\date{{\small Institute of Theoretical Physics Warsaw University \\
ul. Ho\. za 69, 00-681 Warsaw, Poland\\
July 1997} \\
\vspace{-5cm}  \hfill {\bf IFT/9/97} \vspace{5cm} }

\begin{document}

\maketitle

\begin{abstract}

We study the similarity renormalization scheme for hamiltonians 
to the fourth order in perturbation 
theory using a model hamiltonian for fermions coupled to bosons.
We demonstrate that the free finite parts of counterterms can be chosen 
in such a way that the $T$-matrix is covariant up to the fourth order 
and the eigenvalue equation for the physical fermion reduces
to the Dirac equation.
Through this choice, the systematic renormalization scheme reproduces
the model solution originally proposed by G{\l}azek and Perry.

\end{abstract}

\section{Introduction}
Our study of the similarity renormalization scheme 
for hamiltonians \cite{GW1} is done in the model which consists 
of only two sectors in the Fock space. 
This great simplification of the space of states allows complete 
analysis of the renormalization scheme and still includes typical factors 
and divergences that appear in quantum field theory. Our model is 
based on Yukawa theory.

The hamiltonian of Yukawa theory truncated to one 
fermion and one fermion plus one boson Fock sectors leads to 
infinities in the fermion-boson $T$-matrix. 
Therefore, we introduce a cutoff $\La$ for the momentum transfer in 
the interaction part of the hamiltonian. 
The similarity transformation allows us to construct counterterms in the
initial hamiltonian in  such a way that the 
renormalized hamiltonian gives finite and cutoff independent results for 
the $T$-matrix. 
We construct renormalized hamiltonians using expansion in powers of the 
effective fermion-boson coupling constant and including terms up to 
the fourth order.

In the similarity renormalization scheme, one constructs  effective 
hamiltonians $H_\lambda$ which are functions of the width $\lambda$. 
$H_\lambda$ is obtained from the initial hamiltonian $H_\La$ with the imposed 
cutoff $\La$ and added counterterms by a unitary transformation. 
The transformation and counterterms are found order by order 
in perturbation theory 
using the requirement that matrix elements of $H_\lambda$ are independent of 
the cutoff $\La$ when the cutoff goes to infinity. 

To find the unknown finite parts of the counterterms we calculate the 
$T$-matrix for fermion - boson scattering. The condition that the 
$T$-matrix is covariant can be satisfied and it implies relations 
between the finite parts of different counterterms. We also demand, 
that the physical fermion 
is described by the Dirac equation with the fermion mass equal to the 
fermion mass term in the fermion-boson sector.
This demand also provides a relation between the finite parts of counterterms 
and it is called the threshold condition \cite{TD}.

The model hamiltonian we study was originally considered by G{\l}azek 
and Perry \cite{GP}. They guessed the form of counterterms which
remove divergences in $T$-matrix and they obtained covariant 
results for the $T$-matrix to all orders.

Our main question about the model was if the systematic similarity 
calculation, carried out in perturbation theory, produces the same solution to 
the hamiltonian renormalization problem as guessed by G{\l}azek and Perry.
The cutoff in the model is limited by the triviality bound 
\cite{GP} but one can assume that the coupling constant 
is small enough for reliable use of the perturbation theory. 

Section 2 presents the model. Sections 3 and 4 describe its 
renormalization to the
fourth order. Section 5 explains connection with Ref. \cite{GP}
We conclude in Section 6 and Appendix contains key details. 

\section{Model}

The initial hamiltonian is a light-front hamiltonian for Yukawa theory 
projected on two Fock-space sectors: one with a fermion and one 
with a fermion and a boson. 

\begin{equation} 
H_\La=H_{0f}+H_{0fb}+H_Y+H_+ +X_\La \; ,
\end{equation}
where the free part is
\begin{eqnarray}
H_{0f} &=& \sum_\sigma \int [p] \ket{p\sigma}\bra{p \sigma} 
\frac{p^2+m^2}{p^+}\; ,  \label{H0} \\
H_{0fb} &=& \sum_\sigma \int [p,k] 
\ket{p\sigma,k} \bra{p \sigma,k}\left( \frac{p^2+m^2}{p^+}+
\frac{k^2+\mu ^2}{k^+}\right) \;\;\;,
\end{eqnarray}
boson creation and annihilation vertices are
\begin{eqnarray}
H_Y&=&g \sum_{\sigma_1, \sigma_2}\int [p_1,p_2,k] \theta(
\La^2 - M_{p_2,k}^2)
2(2\pi)^3 \delta^3(p_1-p_2-k) \times \nonumber \\
&& \times \left[ \ket{p_2 \sigma_2 ,k}\bra{p_1 \sigma_1}
\bar{u}(p_2,\sigma_2) u(p_1,\sigma_1) + h.c.\right] = 
H_{>\!\!-}+H_{-\!\!<}\;\;\;,
\end{eqnarray}
and the seagull term is
\begin{eqnarray}
H_+&=&g^2 \sum_{\sigma_1, \sigma_2}\int [p_1,p_2,k_1,k_2] 
\theta(\Lambda^2 - {\cal M}_1^2)\theta(\Lambda^2 - {\cal M}_2^2)
2(2\pi)^3 \delta^3
(p_2+k_2-p_1-k_1) \times \nonumber \\
&&
\times \ket{p_2\sigma_2 ,q_2}\bra{ p_1\sigma_1,k_1} \bar{u}(p_2,\sigma_2) 
\frac{\gamma^+}{2(p_1^+ + k_1^+)}u(p_1,\sigma_1) \; .
\end{eqnarray}

$X_{\La}$ in Eq.(1) is an unknown counterterm. We have introduced cutoffs 
on the invariant 
mass ${\cal M}^2=(p+k)^2$ of the two particle sector in the interaction parts 
of the hamiltonian, $H_{Y}$ and $H_+$ (see also \cite{BGP}). 
The integration measure is 
\begin{equation}
[p]=\frac{d^2p^\perp dp^+}{2(2\pi)^3p^+}\;\;,
\end{equation}
\begin{equation}\ket{p}=a_p^\dagger \ket{0} \; \; , \end{equation}
and $\delta^3(p)=\delta^2(p^\perp)\delta(p^+)$. Standard 
light-front commutation relations are
\begin{equation}
[a_p, a_k^\dagger] = 2(2\pi)^3 p^+ \delta^3(p-k) \; .
\end{equation}

\section{Renormalization}

The similarity transformation $S_\lambda$ transforms $H_\La$ to a
band-diagonal hamiltonian $H_\la$,
\begin{equation} 
H_\la=S_\lambda^\dagger H_\La S_\lambda \; .
\end{equation}
Expressions for $S_\lambda$ and $H_\la$ are found in perturbation 
theory \cite{GW1}. 
$X_\La$ in $H_\La$ is fitted order by order in $g$, so that $H_\la$ 
does not have $\La$ dependent (i.e. divergent) matrix elements for 
$\La \rightarrow \infty $. This can be guaranteed 
in any finite order in perturbation theory.

In the second order the transformation gives:
\begin{equation}
H_2^\lambda=f_\la \left(H_++X_2-\frac{1}{2} \left[\underline{(1-f_\la)H_Y},
(1+f_\la)H_Y\right]\right) \; .
\end{equation}
The  underlining denotes the energy denominator and $f_\la$ is the 
diagonal proximum operator (see  Appendix).

In the fermion-boson -- fermion-boson sector Eq. (10) reads
\begin{equation}
H^\lambda_{2fb-fb}=f_\la\left(H_+ -\frac{1}{2}\underline{ 
(1-f_\la) H_{>\!\!-}}(1+f_\la)H_{-\!\!<}
+\frac{1}{2}(1+f_\la)H_{>\!\!-}\underline{(1-f_\la) H_{-\!\!<}}\right)\;\;.
\end{equation}
This expression is not divergent for $\La \rightarrow \infty$, thus no 
counterterm is needed in this sector. However, in the fermion-fermion 
sector, one obtains
\begin{equation}
H^\lambda_{2f-f}=- \underline{ (1-f_\la) H_{-\!\!<}}H_{>\!\!-}+X_{2\La}\;\;. 
\label{loop}
\end{equation}
The loop integration in the first term is linearly divergent. The form of 
this divergence dictates the form of the second order counterterm. 
Explicitly, one has to choose
\begin{equation}
X_{2\La}= \sum_\sigma \int [p] \ket{p\sigma}\bra{p \sigma} \frac{1}{p^+}
\frac{g^2}{16\pi^2}\left[\frac{1}{2}\La^2+(3m^2-\mu^2)\log\frac{\La^2}{m^2}+A
\right] \;, \label{X2}
\end{equation}
where $A$ is an undetermined constant. 

Higher order calculations lead to the following expressions 
for $X_{3\La}$ and $X_{4fb-fb \La}$:
\begin{eqnarray}
X_{3\La}&=&X_{3Y}+X_{3+}=\frac{1}{4}\frac{g^2}{16\pi^2}\log \frac{\La^2}{C}H_Y+
 \nonumber \\ &&
 	+\sum_{\sigma_1, \sigma_2}\int [p_1,p_2,k] 
\theta(\Lambda^2 - {\cal M}_{p_2,k}^2)
	2(2\pi)^3 \delta^3(p_1-p_2-k) \times  \nonumber \\
&&\times \frac {3}{2} \frac{g^3}{16\pi^2}m \, \log \frac{\La^2}{D}
	\left[ \ket{p_2 \sigma_2 ,k}\bra{p_1 \sigma_1}
	\bar{u}(p_2,\sigma_2) \frac{\gamma^+}{2p_1^+}u(p_1,\sigma_1) + h.c.
\right] \; ,
	\\
X_{4fb-fb\La}&=&\frac{1}{2}\frac{g^2}{16\pi^2}\log \frac{\La^2}{B}H_+ \; ,
\label{X4}
\end{eqnarray}
where $B$, $C$ and $D$ are finite unknown constants.

There is also another term of order $g^4$ in the fermion-fermion 
part of $X_\La$.
We did not calculate it because our goal was to investigate the  possibility 
of fitting finite parts of counterterms by requesting the $T$-matrix 
covariance in the fermion-boson channel
(Section 4.1) and the emergence of the Dirac equation for physical fermions.
As $X_{4f-f\La}$ does not contribute either to $T_4$ or the second order Dirac 
equation, it was irrelevant for our considerations. 
Also, $X_{4f-f\La}$  is 
more complicated to calculate than the terms we need to discuss here, 
because of two correlated loop integrations. 

\section{Finite parts of the counterterms}

The renormalization procedure does not determine values of the 
finite parts of counterterms.
To find them we need to introduce extra conditions. 
In principle, the constants should  be fitted to match experiment. It is 
interesting to look for theoretical requirements of symmetries, which may 
constrain these constants.
The $T$-matrix calculated with the general counterterms 
(\ref{X2})-(\ref{X4}) is not automatically
covariant. So, the covariance of the $T$-matrix provides useful conditions. 
Another condition will be provided by requiring that the full Hamiltonian 
eigenvalue equation could be reduced to a free Dirac equation.

\subsection{$T$-matrix}

We calculate our $T$-matrix using the formula
\begin{equation} 
T(E) = H_I + H_I \frac{1}{E-H_0 + i\epsilon} H_I + \cdots \; . \label{T}
\end{equation}
The second order $T$-matrix has a covariant form and does not depend 
on the counterterms. 
$X_\La$ starts contributing in the fourth order. The explicit $\La$ 
dependence of counterterms cancels divergences in the loop 
integrations in other terms. So, $T_4$ is finite. However, it is 
not  covariant automatically.

\begin{displaymath}
\bra{p_2\sigma_2,k_2} T_4 \ket{p_1 \sigma_1, k_1} =
\frac{g^4}{16\pi^2}
\theta(\Lambda^2 - {\cal M}_1^2)\theta(\Lambda^2 - {\cal M}_2^2)
2(2\pi)^3 \delta^3(p_2+k_2-p_1-k_1) \times
\end{displaymath}
\begin{equation} 
\times \left[ \Gamma_1(s)\slsh{P} +\Gamma_2(s) + 
\Gamma_3(s)\frac{\gamma^+}{2(p_1^+ + k_1^+)} 
\right] u(p_1,\sigma_1)\;\;. 
\end{equation}
To obtain a covariant result for $T_4$ we demand that the function 
$\Gamma_3(s)$ vanishes for arbitrary $s$. Its explicit form reads
\begin{equation}
\Gamma_3(s) = \frac{1}{s-m^2}\left[(s-m^2)\frac{1}{2}\log\frac{C}{B} +
 3m^2\log\frac{m^2}{D}-A +16\pi^2\alpha_f(s)(s-m^2)+\gamma_f(s)\right] \; ,
\end{equation}
where  
\begin{equation} 
s= (p_1+k_1)^2 = {\cal M}_1^2 \; ,
\end{equation}
and functions $\alpha_f(s)$ and $\gamma_f(s)$ are given in Appendix.
As $16\pi^2\alpha_f(s)(s-m^2)+\gamma_f(s)$ turns out to be real and independent of 
$s$, the condition $\Gamma_3(s)=0$ implies two relations:
\begin{equation}
B=C  \label{BC}
\end{equation}
and
\begin{equation}
A=-m^2+\mu^2\log\frac{\mu^2}{m^2}+3m^2\log\frac{m^2}{D} \; . \label{A}
\end{equation}

\subsection{Dirac equation}

To describe a physical state in terms of free Fock states one considers 
the eigenvalue equation
\begin{equation}
H_\La \ket{P \sigma}_{physical}=\frac{P^{\perp 2}+m^2}{P^+}
\ket{P \sigma}_{physical} \;\;. \label{Dir}
\end{equation}
The physical fermion state is a superposition of the bare fermion and 
fermion-boson states:
\begin{equation}
\ket{P\sigma}_{physical}=\sum_{\sigma_2}c^\sigma_{\sigma_2}\ket{P\sigma_2}+
\sum_{\sigma_2}\int[p,k]2(2\pi)^3\delta^3(P-p-k)
\phi^\sigma_{\sigma_2}(x,M^2)\ket{p \sigma_2,k} \;\;.
\end{equation}

By following steps from ref. \cite{GP} one can reduce Eq.(\ref{Dir}) to 
\begin{equation}
\left( \Xi_1 \slsh P _m - \Xi_2 m + 
\Xi_3 \frac{\gamma^+}{2P^+}\right)\psi=0\;\;,
\end{equation}
for the one-body sector wavefunction $\psi$. Using our hamiltonian with 
counterterms restricted by conditions (\ref{BC})-(\ref{A}), one gets
\begin{eqnarray}
\Xi_1 &=& 1+\frac{g^2}{16\pi^2} \left[ \frac{3}{2} \log\frac{\La}{D} - 
\beta(m^2)\right] +o(g^4)\; , \label{xi1}\\
\Xi_2 &=& 1 +\frac{g^2}{16\pi^2}\alpha(m^2)+o(g^4) \label{xi2}\; ,\\
\Xi_3 &=& 0 +o(g^4)\;.\label{xi3}
\end{eqnarray}
Our earlier demand of the $T$-matrix covariance established the value 
of the mass counterterm $X_2$ in a way that also leads to the vanishing 
of $\Xi_3$ in order $g^2$. 

In general, one can expand both non-zero $\Xi$'s in a power series in $g$
\begin{equation}
\Xi=\Xi^{(0)}+\Xi^{(2)}g^2+\Xi^{(4)}g^4+\cdots
\end{equation}
and one can translate the requirement that $m$ is the mass of 
physical fermions, 
\begin{equation}
\left( \slsh P_m-\frac{\Xi_2^{(0)}+\Xi_2^{(2)}g^2+\Xi_2^{(4)}g^4+\cdots}{
\Xi_1^{(0)}+\Xi_1^{(2)} g^2+\Xi_1^{(4)}g^4+\cdots} m\right)\psi=0 \; ,
\end{equation}
into the condition for all coefficients
\begin{equation}
\Xi_1^{(i)} = \Xi_2^{(i)} \; . \label{Xi}
\end{equation}
This is the threshold condition which makes the $T$-matrix threshold 
to appear at $s=(m+\mu)^2$, where $m$ is the position of its fermion pole.

Let us investigate which terms of $H$ contribute to $\Xi^{(i)}$. 
If one puts $g=0$ then, the only condition one gets is 
\begin{equation}
\ket{P\sigma}_{physical}=\ket{P\sigma}\;\;.
\end{equation}
Technically, the zeroth order terms $\Xi_1^{(0)}$ 
and $\Xi_2^{(0)}$ come from the 
inversion of $\sum_\sigma u_{P\sigma m} \bar{u}_{P\sigma m}=
{\not\!\slsh P}_m +m$, which is a part of $H_{>\!\!-}H_{-\!\!<}$. 
Dirac equation results in this order automatically; $\Xi_1^{(0)} 
= \Xi_2^{(0)}$.

One can easily see that the second order terms $\Xi_1^{(2)}$ 
and $\Xi_2^{(2)}$ 
partly come from the term $H_{>\!\!-}X_{+3}$. So, one 
needs third order vertex 
corrections, such as $X_{+3}$,
to know all second order contributions to the Dirac 
equation. There is an unknown finite parameter $D$ in $X_{+3}$. The 
condition $\Xi_1^{(2)}=\Xi_2^{(2)}$ and Eqs. 
(\ref{xi1})-(\ref{xi2}) lead to 
\begin{equation}
\log \frac{D}{m^2}=\frac{2}{3}\cdot 16 \pi^2 \left[\alpha_f(m^2) +
\beta_f(m^2) \right] \;\;. \label{D}
\end{equation} The functions 
$\alpha_f(s)$ and $\beta_f(s)$ are given in Appendix.

We see that the requirement that $m$ is equal to the mass of physical 
fermions implies one more condition on the free parts of counterterms.

\subsection{Discussion}

Collecting conditions (\ref{BC}), (\ref{A}) and (\ref{D}) together, 
and looking at the structure 
of the counterterms, we can observe the following.
$X_{3Y}$ can be accounted for by changing the  coupling constant 
of $H_Y$
\begin{equation}
g \rightarrow g+ \frac{g^3}{64\pi^2}\log \frac{\La^2}{C} \label{g}
\end{equation}
in the original hamiltonian, 
while  $X_{4fb-fb \La}$ shifts $g^2$ in the seagull 
term $H_+$:
\begin{equation}
g^2 \rightarrow g^2+ \frac{g^4}{32\pi^2}\log 
\frac{\La^2}{C}\;\;.
\end{equation}
So, these two counterterms can be absorbed in one, $\La$-dependent  
coupling constant (\ref{g}). We have to stress that, in physical results, 
$\La$ dependent logarithms $\log \frac{\La^2}{m^2}$ cancel out, leaving 
\begin{equation}
g+\frac{1}{4}\frac{g^3}{16\pi^2}\log\frac{m^2}{C}\;\;. \label{gC}
\end{equation}
Thus, $g$ and $C$ will never appear independently, and we have one 
parameter, combination (\ref{gC}), that can be fixed from experiment.

$X_2$ shifts the mass in the one fermion free energy. 
Sum of $H_Y$ and one of the third order counterterms, $X_{3+}$, 
reproduces the same $\bar{u}u$ coupling but with shifted mass 
of the spinor in the one-particle sector, according to the formula
\begin{equation}
\left( 1+\frac{\gamma^+\delta m}{2p^+} \right) u_m(p,\sigma)=
u_{m+\delta m}(p,\sigma) \; . \label{ver}
\end{equation}
\section{ Comparison with Ref. [3]}

It was shown in Ref. \cite{GP} that, in this model, 
to get finite and covariant results for the 
$T$-matrix to all orders of perturbation theory, and to get the mass in the 
Dirac equation which is required by the threshold condition, it is enough to 
(1) add to the bare cut-off hamiltonian a term that shifts the mass of 
fermions in the free part $H_{0f}$,
(2) correspondingly, change the spinor mass in the vertex, see Eq. 
(\ref{ver}),
and (3) allow the coupling to depend on $\La$.

When one rewrites the hamiltonian of Ref. \cite{GP} using 
the invariant mass cutoff and expands it in powers of $\tilde g(m^2)$ 
up to the fourth order, one gets the same result as obtained in 
our similarity calculation with 
\begin{equation} 
g+\frac{1}{4}\frac{g^3}{16\pi^2}\log \frac{m^2}{C} 
\end{equation}
replaced by
\begin{equation}
\tilde g(m)-\frac{1}{2}\tilde g^3(m)\alpha_f(m^2) \;.
\end{equation}
\noindent So, one can choose $C$ leading to the same result 
as in Ref. \cite{GP}. 

\section{Conclusion}

This work provides an example of application of the similarity renormalization 
scheme in its algebraical version. We have shown
how this systematic 
procedure leads from a divergent hamiltonian to a finite one. 
The finite hamiltonian gives a covariant scattering matrix in perturbation 
theory. 

The hamiltonian we used was known to lead to covariant results when one 
introduced special counterterms. The question was if 
a systematic  procedure, the similarity renormalization scheme, would
 produce the same solution. The answer is yes. 

On the other hand, Ref. \cite{BGP} 
has recently suggested that the model may find 
applications in pion-nucleon physics when another Fock sector, with one 
fermion and two bosons, is included. Therefore, our work also suggests that a 
systematic improvement in the light-front hamiltonian approach to relativistic
nuclear physics may be achivable using the similarity renormalization group 
techniques. 

\section*{Acknowledgment}

This research  has been supported in part by Maria Sk{\l}odowska-Curie 
Foundation under Grant No. MEN/NSF-94-190. The authors would like to 
thank Stan G{\l}azek for many discussions. 

\appendix
\section*{Appendix}

For any operator $A$,
\begin{equation}
A = \int\ket{1}\bra{2}A_{12} \; ,
\end{equation}
$\underline{A}$ is defined as 
\begin{equation}
\underline{A} = \int\ket{1}\bra{2}\frac{1}{E_2- E_1}A_{12} \; ,
\end{equation}
where $E$'s are eigenvalue of $H_0$. $\underline{A}$ is a solution of an equation:
\begin{equation}
[\underline{A},H_0]=A \; .
\end{equation}

Action of diagonal proximum operator $f_\la$ is defined as follows :
\begin{equation}
f_\la A = \int\ket{1}\bra{2} \tilde{f}_\la(1,2) A_{12}\; .
\end{equation}
We have chosen 
\begin{equation}
\tilde{f}_\la(1,2) = \theta( \la^2 - | {\cal M}_1^2 - {\cal M}_2^2| ) \; .
\end{equation}
\bigskip

Functions $\alpha(s)$, $\beta(s)$ and $\gamma(s)$ are defined by 
\begin{eqnarray*}
\alpha(s)&=&-\frac{1}{16\pi^2}\int dM^2 \,dx\; \theta(\Lambda^2 - M^2) 
\; \frac{x}{M^2-s + i \epsilon } \; ,\\ 
\beta(s)&=&-\frac{1}{16\pi^2}\int dM^2 \,dx \; \theta(\Lambda^2 - M^2) 
\; \frac{1}{M^2-s + i \epsilon }\; ,\\
\gamma(s)&=&\int dM^2 \, dx\; \theta(\La^2-M^2) 
\; \frac{(1-x)M^2-\mu^2+(1-x)m^2}{M^2-s + i \epsilon } \; ,
\end{eqnarray*}
where $x$ is integrated over the whole kinematically allowed 
region. Their finite parts are defined by
\begin{eqnarray*}
\alpha_f(s) &=&\lim_{\La\rightarrow\infty}\left[\alpha(s)+
\frac{1}{2\cdot16\pi^2}\log\frac{\La^2}{m^2}\right] \; ,\\
\beta_f(s)  &=&\lim_{\La\rightarrow\infty}\left[\beta(s)+
\frac{1}{16\pi^2}\log\frac{\La^2}{m^2}\right] \; ,\\
\gamma_f(s) &=&\lim_{\La\rightarrow\infty}\left[ \gamma(s) 
- \frac{1}{2} \La^2 -\frac{1}{2}(s-m^2-2\mu^2)\log \frac{\La^2}{m^2} 
\right] \; .
\end{eqnarray*}

\end {document}